\documentclass[twocolumn,aps,prl,amsmath,showpacs]{revtex4-1}
\usepackage{graphicx,dcolumn,bm,amssymb,amsmath}
\usepackage{color}

\usepackage{float}

\usepackage{hyperref}
\def\be{\begin{equation}}
\def\ee{\end{equation}}

\begin{document}
\title{\Large Low-energy optical phonons induce glassy-like vibrational and thermal anomalies in ordered crystals}
\author{Matteo Baggioli$^{1,2}$ and  Alessio Zaccone$^{3,4,5}$}
\affiliation{${}^1$Instituto de Fisica Teorica UAM/CSIC, c/Nicolas Cabrera 13-15,
Universidad Autonoma de Madrid, Cantoblanco, 28049 Madrid, Spain.}
\affiliation{${}^2$Crete Center for Theoretical Physics, Institute for Theoretical and Computational Physics,
Department of Physics, University of Crete, 71003 Heraklion, Greece.}
\affiliation{${}^3$Department of Physics ``A. Pontremoli", University of Milan, via Celoria 16, 20133 Milan, Italy}
\affiliation{${}^4$Department of Chemical Engineering and Biotechnology,
University of Cambridge, Philippa Fawcett Drive, CB30AS Cambridge, U.K.}
\affiliation{${}^5$Cavendish Laboratory, University of Cambridge, JJ Thomson
Avenue, CB30HE Cambridge, U.K.}

\begin{abstract}
\noindent 
It is widely accepted that structural glasses and disordered crystals exhibit anomalies in the their thermal, mechanical and acoustic properties as manifestations of the breakdown of the long-wavelength approximation in a disordered dissipative environment. However, the same type of glassy-like anomalies (i.e. boson peak in the vibrational density of states (VDOS) above the Debye level, peak in the normalized specific heat at $T\simeq10 K$ etc) have been recently observed also in perfectly ordered crystals, including thermoelectric compounds.
Here we present a theory that predicts these surprising effects in perfectly ordered crystals as a result of low-lying (soft) optical phonons. In particular, it is seen that a strong boson peak anomaly (low-energy excess of modes) in the VDOS can be due almost entirely to the presence of low-energy optical phonons, provided that their energy is comparable to that of the acoustic modes at the Brillouin zone boundary. The boson peak is predicted also to occur in the heat capacity at low $T$. In presence of strong damping (which might be due to anharmonicities in the ordered crystals), these optical phonons contribute to the low-$T$ deviation from Debye's $T^{3}$ law, producing a linear-in-$T$ behavior which is typical of glasses, even though no assumptions of disorder whatsoever are made in the model. These findings are relevant for understanding and tuning thermal transport properties of thermoelectric compounds, and possibly for the enhancement of electron-phonon superconductivity. 
\end{abstract}


\maketitle


\section{Introduction}
Structural glasses as well as disordered crystals are known to present anomalies in their vibrational spectra and in their low-T properties, such as the specific heat and the thermal conductivity. The most studied effect is the so-called boson peak in the vibrational density of states (VDOS), which manifests itself as a peak (excess of modes) at THz frequencies in the VDOS normalized by the Debye law $\omega^{2}$~\cite{Buchenau}. The boson peak is also responsible for deviations from the Debye $T^{3}$ law in the specific heat, including a peak at temperatures on the order of $10$K and for a low-T plateau in the thermal conductivity.

Several theories have been proposed to explain these effects, starting from heterogeneous-elasticity theory~\cite{Schirmacher}, and include such diverse approaches as: randomly-distributed soft anharmonic modes~\cite{Gurevich,Parshin}, local inversion-symmetry breaking connected with nonaffine deformations~\cite{Zaccone2011, Zaccone2013, Milkus}, phonon-saddle transition in the energy landscape~\cite{Parisi}, density fluctuations of arrested glass structures~\cite{Goetze}, and broadening/lowering of the lowest van Hove singularity in the corresponding reference crystal due to the distribution of force constants~\cite{Diezemann,Elliott}.
Theories of the specific heat anomaly at low-$T$ have been based mostly on the two-level system (TLS) mechanism~\cite{Anderson}. More recently, a new approach has shown that the boson peak anomaly arises from the competition between acoustic propagating phonons and viscous damping, thus providing a more general explanation of the origin of these anomalies, which applies to both glasses and crystals~\cite{Baggioli}.

Recent experimental works have demonstrated the existence of the very same anomalies (boson peak and heat capacity anomaly) also in perfectly ordered crystals, such as molecular single crystals~\cite{Tamarit,Pardo2,Jezowski} and non-centrosymmetric perfect crystals such as $\alpha$-quartz~\cite{Monaco}.

Furthermore, the same phenomenology of a pronounced boson peak in the VDOS has been observed in thermoelectric crystals~\cite{Suekuni}. Thermoelectric crystals can be used to convert heat fluxes into charge carrier fluxes, and the efficiency is high when thermal conductivity is low and at the same time electrical conductivity is high. Low thermal conductivity can be achieved with guest (caged) heavier atoms that undergo low-energy vibrations, thus inducing low-energy optical modes~\cite{Tse, Christensen}. It is important to properly understand the link between these low-energy vibrations of the caged atoms and the thermal conductivity, especially since the caged atoms induce a strong boson peak in the VDOS, and the thermal conductivity is, in turn, expressed as an integral involving the VDOS.

These observations call for a deeper understanding of the fundamental origin of these anomalies, well beyond the paradigms developed for glasses, which fail to explain these observations in fully ordered crystals, for  obvious reasons.

Here we reexamine this problem from the point of view of the competition between propagating phonon excitations and quasi-localized diffusive excitations produced by viscous damping, and, crucially, we account for the role of both acoustic phonons and optical phonons.
In particular, the theory predicts that, in systems with soft (low-energy) optical phonons piling up at low energy, the boson peak is entirely controlled by the low-energy optical phonons. Similar behaviours as that predicted by our theory has been reported in contemporary experimental work. 

In particular,~\cite{Moratalla} has reported experimental observations of a strong boson peak in perfectly ordered crystals of halomethanes due to low-lying optical modes, whereas ~\cite{Mori1,Mori2} experimentally observed the upturn of the VDOS predicted here in organic molecular systems with high degree of crystallinity (and presumably very soft optical modes) such as starch and glucose. 
Similar behaviors have been measured also in thermoelectric crystals~\cite{Taka,Mori3} where a boson peak in the VDOS is also observed, and where, interestingly, low-energy vibrations (so-called rattling) of caged compounds give rise to an upturn in the VDOS at vibrational energies below the boson peak~\cite{Suekuni}.

\section{Theory}
\subsection{Acoustic phonons}
The starting point is to relate the VDOS to the Green's functions of the vibrational degrees of freedom of the system. In particular, if we consider just acoustic, longitudinal (L) and transverse (T), phonons, the VDOS is given in terms of the Green functions by the following relations already used in ~\cite{Baggioli}:
\begin{equation}
g_{A}(\omega)=-\frac{2\,\omega}{3\,\pi
\,\mathcal{N}}\sum_{q<q_D}\,\text{Im}\left\{2\,G_{TA}(q,\omega)+G_{LA}(q,\omega)\right\},\label{eq1}
\end{equation}.\label{eq1}
which can be derived using the Plemelj identity (see Supplementary Info of ~\cite{Baggioli} for derivation).

The Green's functions for the acoustic phonons can be derived from the standard anharmonic Hamiltonian with cubic and quartic terms in the Taylor expansion for the potential energy and are given by:
\begin{equation}
G_{L,T}(q,\omega)=\frac{1}{c^2_{L,T}q^2-\omega^2-i\,\omega\,\Gamma},\label{eq2}
\end{equation}
where
\begin{align}
&c_{T}^2=\frac{\mu}{\rho}\,,\quad c_{L}^2=\frac{K\,+\,\frac{2\,(d-1)}{d}\,\mu}{\rho}\\
&\Gamma_{TA}=D_{TA} q^{2}\,,\quad \Gamma_{LA}=D_{LA} q^{2}.
\end{align}
Here $\mu$ is the shear elastic modulus, $K$ the bulk modulus, $\rho$ the density. Furthermore, $\Gamma_{TA}$ is the phonon damping for transverse acoustic phonons, and $D_{TA}$ is its corresponding (constant) diffusion coefficient (and respectively for the longitudinal components, with subscript $_{LA}$). \\

As discussed in Ref.~\cite{Baggioli}, the boson peak coincides with the Ioffe-Regel crossover of the transverse phonons, where the onset of diffusive-like modes is controlled by the phonon damping~\cite{Landau-Lifshitz}, which in turn results from anharmonicity~\cite{Rumer}.

\begin{figure}
\centering
\includegraphics[width=7.5cm]{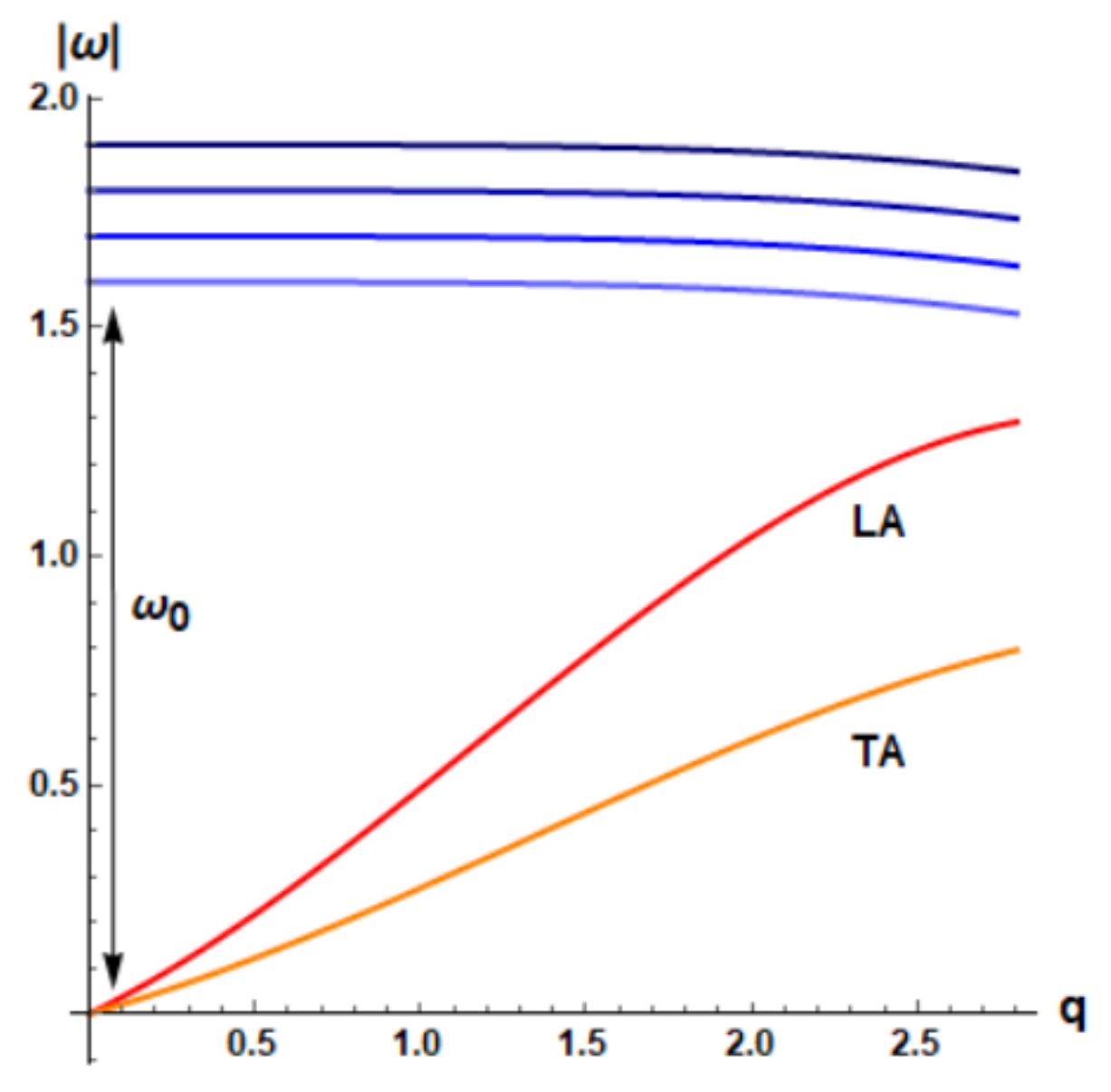}
\caption{Schematic of the model dispersion relations used in the calculations of the VDOS and of the specific heat. Three optical phonons are used in the calculations, which are given by the dispersion relation Eq.\eqref{disp}, and for simplicity are taken to be equally spaced along the $\omega$ axis.}
\label{fig1}
\end{figure}

\subsection{Optical phonons}
Now let us consider the presence of an optical phonon which, e.g. in the case of a crystal with two atoms in the primitive cell, is described at low wavevector $q$ by~(see e.g. \cite{Kosevich}):
\begin{equation}
\omega\,=\,\omega_0\,-\,A\,q^2\,-\,i\,\Gamma_{O}.\label{disp}
\end{equation}
where $\Gamma_{O}$ is the damping contribution for the optical mode. Importantly, $\omega_0$ coincides with the energy gap $\Delta$ of the optical mode, and hence with the mass in a Klein-Gordon scalar field theory.

We assume that also the optical phonon is described by a Damped Harmonic Oscillator (DHO) Green's function of the same form used for the acoustic phonons.
The optical modes thus produce an additional contribution to the VDOS:
\begin{equation}
g_O(\omega)=-\frac{2\,\omega}{3\,\pi
\,\mathcal{N}}\sum_{q<q_D}\text{Im}\left\{G_O(q,\omega)\right\},\label{eq6}
\end{equation}
where the corresponding Green function is given by:
\begin{equation}
G_O(q,\omega)\,=\,\sum_{k=1}^{N_{O}}\frac{1}{-\omega^2\,-\,i\,\Gamma_{O}\,\omega+\omega_{0,k}^2\,+\,A^2\,q^4},\label{eq7}
\end{equation}
where $N_{O}$ is the number of optical modes. For simplicity we keep the same damping for all modes, and we further specify that also the damping of optical modes is diffusive-like, with $\Gamma_{O}=D_{O}q^{2}$. The latter is another assumption of the model which can be justified based on hydrodynamics. Since we are focusing on the low-$q$ behavior of interest here, the diffusive damping $\Gamma_{O}=-i D_{O}q^{2}$ allows us to recover the form of damping prescribed by hydrodynamics~\cite{Parodi} in the $q\rightarrow 0$ limit. In any case, also for optical phonons the root cause of damping is anharmonicity, specifically in the form of decay processes of an optical phonon into two acoustic phonons~\cite{Klemens}.

In the following we study the full VDOS given by:
\begin{equation}
g(\omega)=-\frac{2\,\omega}{3\,\pi
\,\mathcal{N}}\sum_{q<q_D}\,\text{Im}\left\{G_{A}(q,\omega)+G_O(q,\omega)\right\},\label{eq8}
\end{equation}
where $G_{A}=2\,G_{TA}(q,\omega)+G_{LA}(q,\omega)$.
The VDOS thus includes: (i) the acoustic phonons (transverse and longitudinal); (ii) the effect of the optical phonons described above and given by Eqs.\eqref{eq6}-\eqref{eq7}; (iii) the effect of mode damping (which is due to anharmonicity~\cite{Rumer} and can also be related to the viscosity of the solid~\cite{Landau-Lifshitz,visc}).
\begin{figure}
\includegraphics[width=8.5cm]{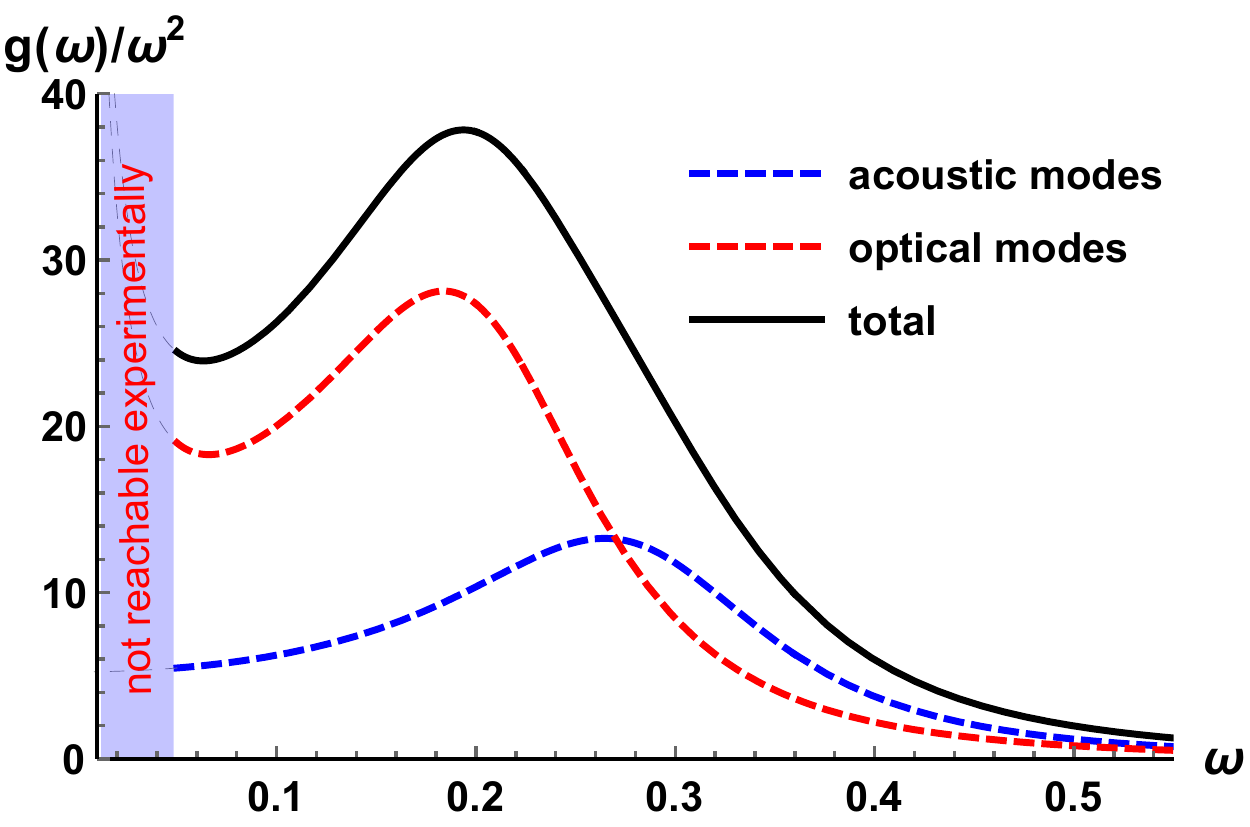}
\centering

\vspace{0.5cm}

\includegraphics[width=8.5cm]{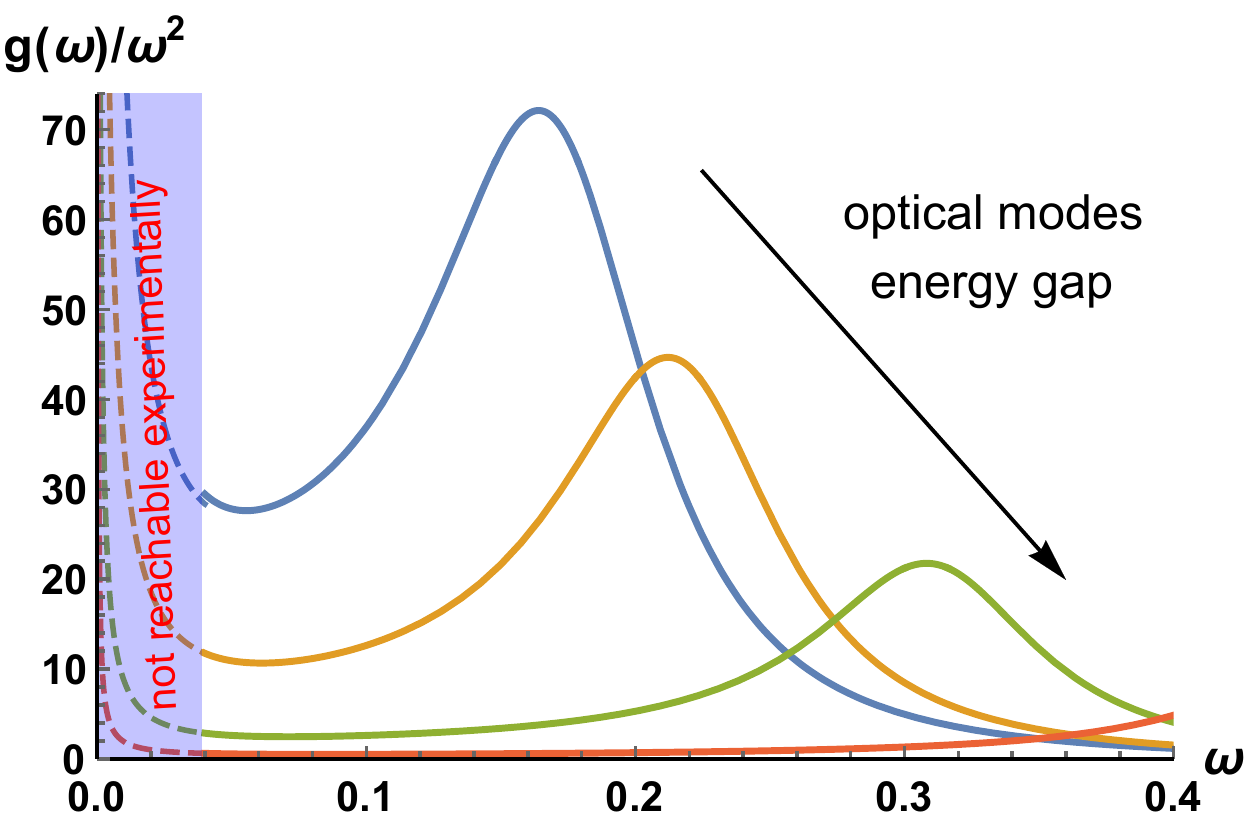}
\caption{\textbf{Top panel:} Acoustic Eq.(1) and optical Eq.(6) contributions to the total VDOS, Eq.\eqref{eq8}. The parameters used in the plot are $A=0.001, N_O=3,D_{O}=0.02$, for the optical modes, and $\mathcal{N}=1, c_{T}=0.47, D_{T}=0.4, c_{L}=0.6, D_{L}=0.5, k_{D}=1$. \textbf{Bottom: } Optical contribution Eq.\eqref{eq6} shown in the top panel as a function of the energy gap $\omega_0$.}

\label{fig2}
\end{figure}
\subsection{Temperature dependence of the damping coefficients}
The damping coefficients $D$ are normally temeperature-dependent. This temperature dependence is different for acoustic and for optical phonons. 
For acoustic phonons, since we are interested in the regime of low wavevector $q$, we assume a temperature dependence of the Landau-Rumer type~\cite{Rumer}, i.e. 
\begin{equation}
D_{LA}(T)=D_{TA}(T) \sim T^{4},\label{eq9}
\end{equation}
which arises from the evaluation of scattering matrix elements of phonon-phonon processes. However, we have checked that this dependence does not introduce significant qualitative changes with respect to choosing temperature-independent coefficients. The same $\sim T^{4}$ dependence arises from Rayleigh scattering due to isolated defects. 
Since however the above temperature-dependence effectively brings the acoustic damping to zero at $T=0$, in order to avoid issues with the evaluation of the Green's function Eq.\eqref{eq2} it is necessary to introduce a small imaginary term in the denominator, $i\epsilon$, which allows us to recover the perfectly harmonic behaviour at $T=0$.

For the optical phonons, it is usually assumed that the anharmonic damping of a low-energy optical phonon results in the decay of the optical phonon into two acoustic phonons, a mechanism proposed originally by Klemens~\cite{Klemens}. From a master equation describing these phonon-phonon processes, one obtains the following Klemens formula for the damping coefficient of optical phonons:
\begin{equation}
D_{O}(T)=D_0\,[1+2/(e^{\beta\hbar\omega/2}-1)]
\end{equation}
Importantly, due to the presence of a Bose-Einstein coefficient, this expression also affects the final calculation for the VDOS. This is not a negligible effect: since the bosons occupation probability increases upon decreasing the normal mode energy $\hbar\omega$, this effect contributes to the upturn in the normalized VDOS at $\omega \rightarrow 0$, as will be shown below. 

\section{Results}
The evaluation of the above model is plotted in Fig.\ref{fig2} for a single low-lying optical phonon. In the top panel, the VDOS clearly displays a strong boson peak which is controlled by the low-energy optical phonon, whereas the acoustic phonons give a much smaller contribution to the boson peak. 
To be precise, as discussed in Ref.~\cite{Baggioli}, the boson peak arises from the competition between phonon propagation and the diffusive-like damped modes. This competition gives rise to a Ioffe-Regel crossover in correspondence to the boson peak. 
In this case, the boson peak is given by the same competition mechanism, but this time the competition is between the low-energy optical modes and the diffusive-like damped modes, whereas the acoustic modes play a minimal role. 
\begin{figure}
\centering
\includegraphics[width=8.5cm]{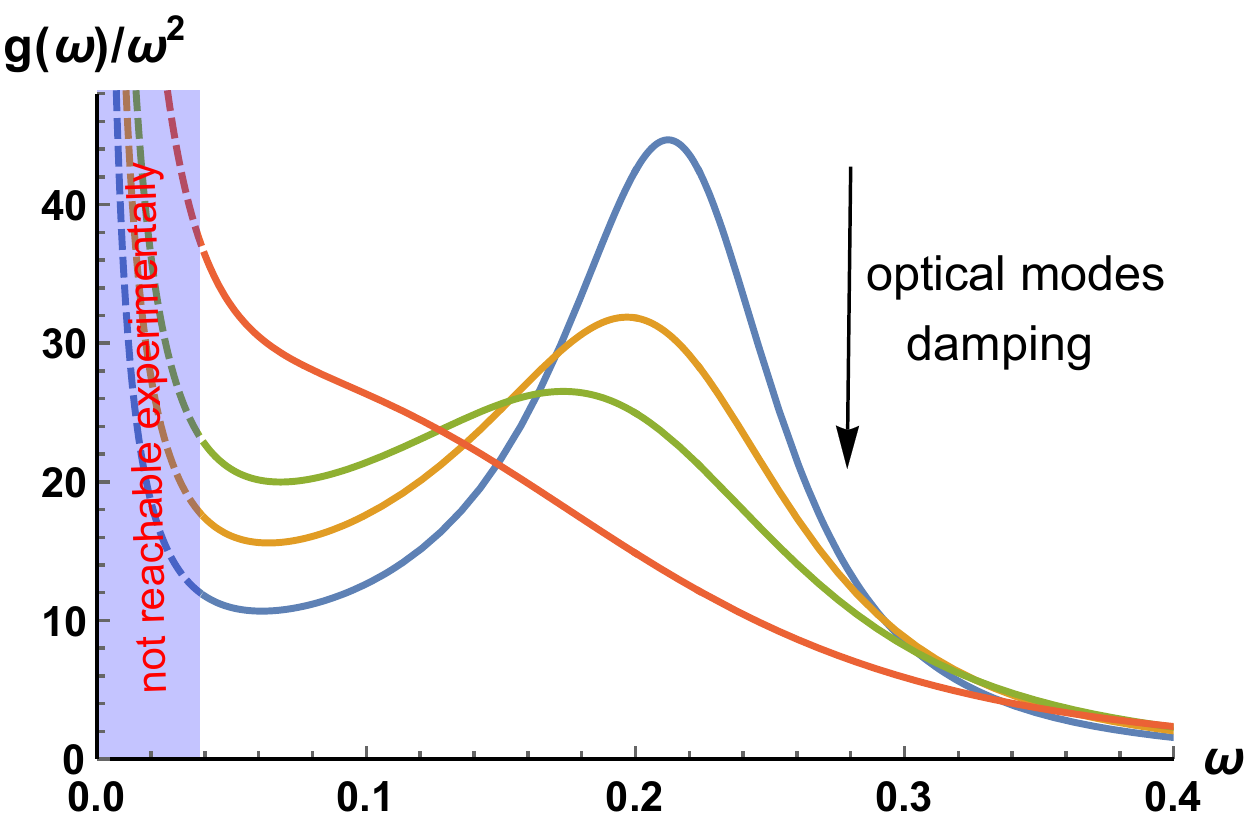}

\caption{Evolution of the optical contribution to the VDOS upon varying the magnitude $D_0$ of the damping coefficient for the optical phonon. In the calculation we took $\hbar=1$ and $k_{B}=1$.}
\label{fig3}
\end{figure}\vspace{0.5cm}

In Fig.\ref{fig2}, bottom panel, the dependence of the optical contribution to the boson peak is shown for different values of the frequency at the Brillouin zone center, $\omega_{0}$, which determines the energy scale of the optical phonon. It is clear that the effect arises only if the optical phonon energy $\omega_{0}$ is sufficiently low, i.e. comparable to the energy of the acoustic phonons at the Brillouin zone boundary. The effect becomes more prominent as $\omega_{0}$ decreases, whereas it vanishes in the limit of large optical gap $\omega_{0}$. This observation allows us to conclude that only low-energy optical phonons can give rise to glassy-like vibrational anomalies in perfectly ordered crystals. 

Also, we note that the low-energy optical phonon is responsible for the upturn of the VDOS upon approaching 
$\omega=0$, an effect which is amplified by the frequency-dependence of the damping coefficient due to the presence of the Bose-Einstein occupation factor in the Klemens formula, Eq. (10). To our knowledge, this is another feature that has never been predicted before. 

The experimental observation of this low-$\omega$ upturn in the normalized VDOS may be difficult, due to some instrumental function or quasi-elastic signal which always hides the VDOS near $\omega=0$. However, this upturn has been clearly observed in recent experiments \cite{Mori1,Mori2} using Raman scattering and terahertz spectroscopy on starch and glucose vitreous systems, which possess a significant degree of crystallinity and where low-energy optical phonons may play a role. Furthermore, optical phonons in such systems of large organic molecules, are expected to occur at low energy and to be strongly damped. 

Importantly, the same low-$\omega$ upturn has been clearly observed also in thermoelectric crystals where caged atoms undergo low-energy ''large" vibrational motions, thus giving rise to low-lying optical modes~\cite{Suekuni}. The inelastic neutron scattering (INS) results of ~\cite{Suekuni} on tetrahedrite crystals (Fig. 4 in that article) appear indeed to fully support the theoretical scenario presented here and which underlies the behaviour in Fig.\ref{fig2}. 

In Fig.\ref{fig3}, the effect of the damping coefficient on the optical contribution to the VDOS is shown. It is evident that, upon increasing the damping, the peak is redshifted, i.e. it moves towards lower frequencies, which is expected in view of the competition mechanism between propagating and diffusive transport. However, contrary to what happens for acoustic phonons, the peak at intermediate frequencies tends to get lower with increasing damping. The spectral weight is transferred towards lower and lower energies producing a strong upturn in the normalized density of states. As we will see later, this is just the manifestation that for large damping the VDOS displays a wide constant plateau $g(\omega)\sim const.$ which moves towards lower frequencies upon increasing the diffusion constant of the phonons. Physically, this feature can be connected with the more and more localized nature of the vibrational degrees of freedom.
\begin{figure}
\includegraphics[width=8.5cm]{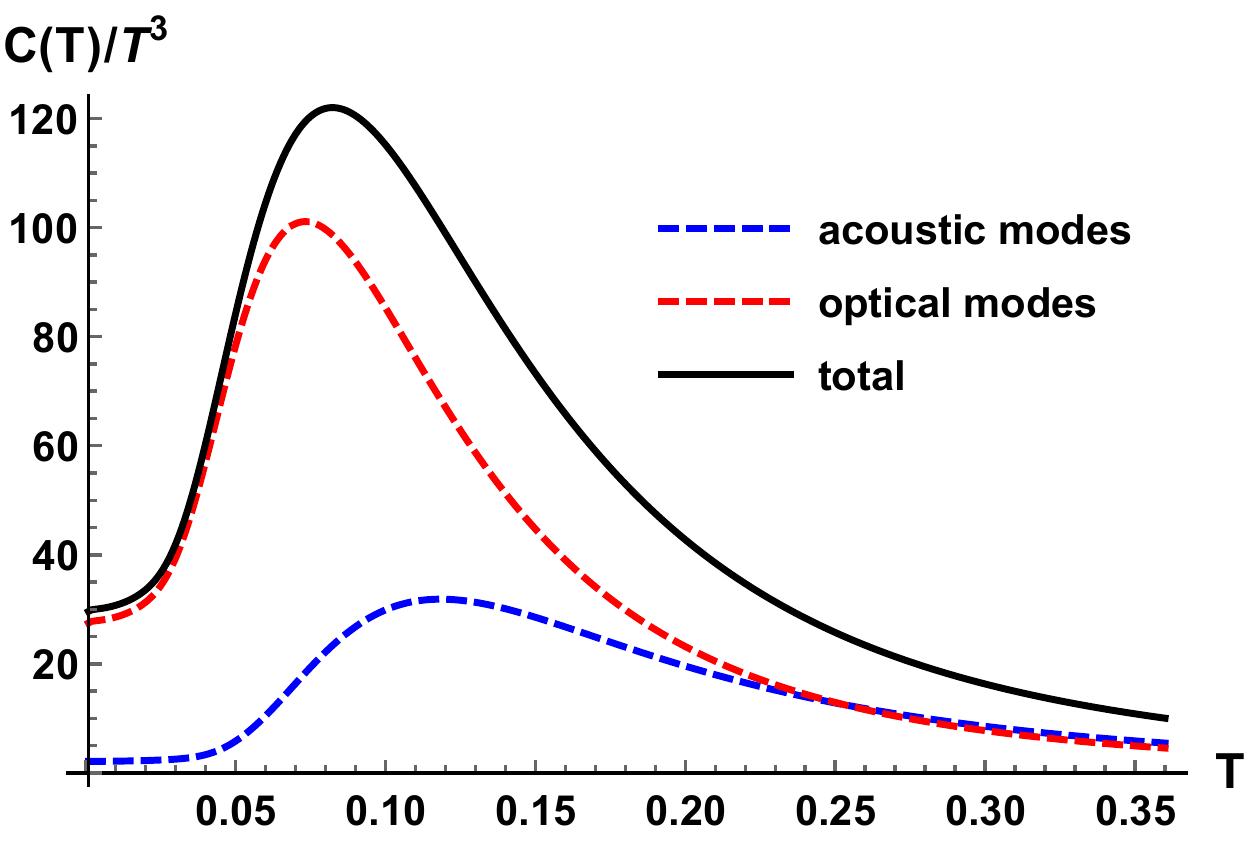}
\centering

\vspace{0.5cm}

\includegraphics[width=8.5cm]{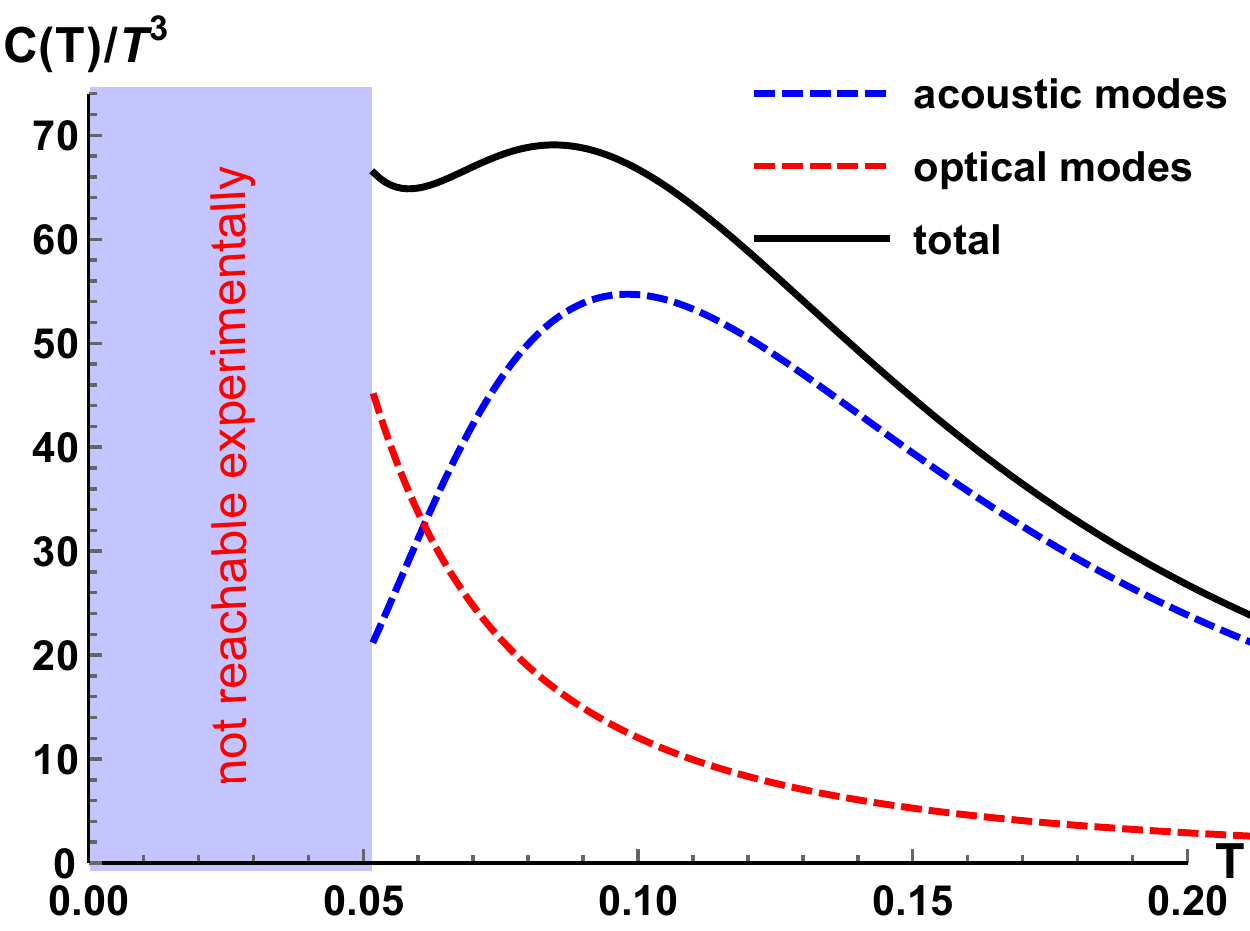}
\caption{\textbf{Top panel:} Specific heat calculated with Eq.\eqref{eq9} on the basis of the VDOS (same as in Fig.\ref{fig2} top panel) given by Eq.\eqref{eq8}: total (black), optical part (red) and acoustic part (blue). In the calculation we took $\hbar=1$ and $k_{B}=1$. \textbf{Bottom panel:} The same specific heat calculated for a much higher value of the optical modes damping. See fig.\ref{fig3} for a zoom of this behaviour and section \ref{TLS} for a discussion on this point.}
\label{fig4}
\end{figure}

\subsection{Specific heat}
With the VDOS obtained in this way, the specific heat can be obtained in the standard way via~\cite{Khomskii}:
\begin{equation}
C(T)\,=\,k_B\,\int_0^\infty \,\left(\frac{\hbar\omega}{2\,k_B\,T}\right)^2\,\sinh \left(\frac{\hbar\omega}{2\,k_B\,T}\right)^{-2}\,g(\omega)\,d\omega.
\end{equation}
Model predictions are shown in Fig. 4 for the two different sets of parameters used in the VDOS calculations.
In the top panel, we see that the optical contribution to the peak in the specific heat is the dominant one, and is much larger than the acoustic contribution. This prediction has been experimentally and numerically confirmed in the work of Moratalla et al. ~\cite{Moratalla}, where DFT calculations of the two contributions were reported, which are in perfect agreement with our Fig. 4 (top panel).  
Also, we see that in this case the specific heat behaves like the Debye prediction $\sim T^{3}$ at low T.
\begin{figure}
\includegraphics[width=8.5cm]{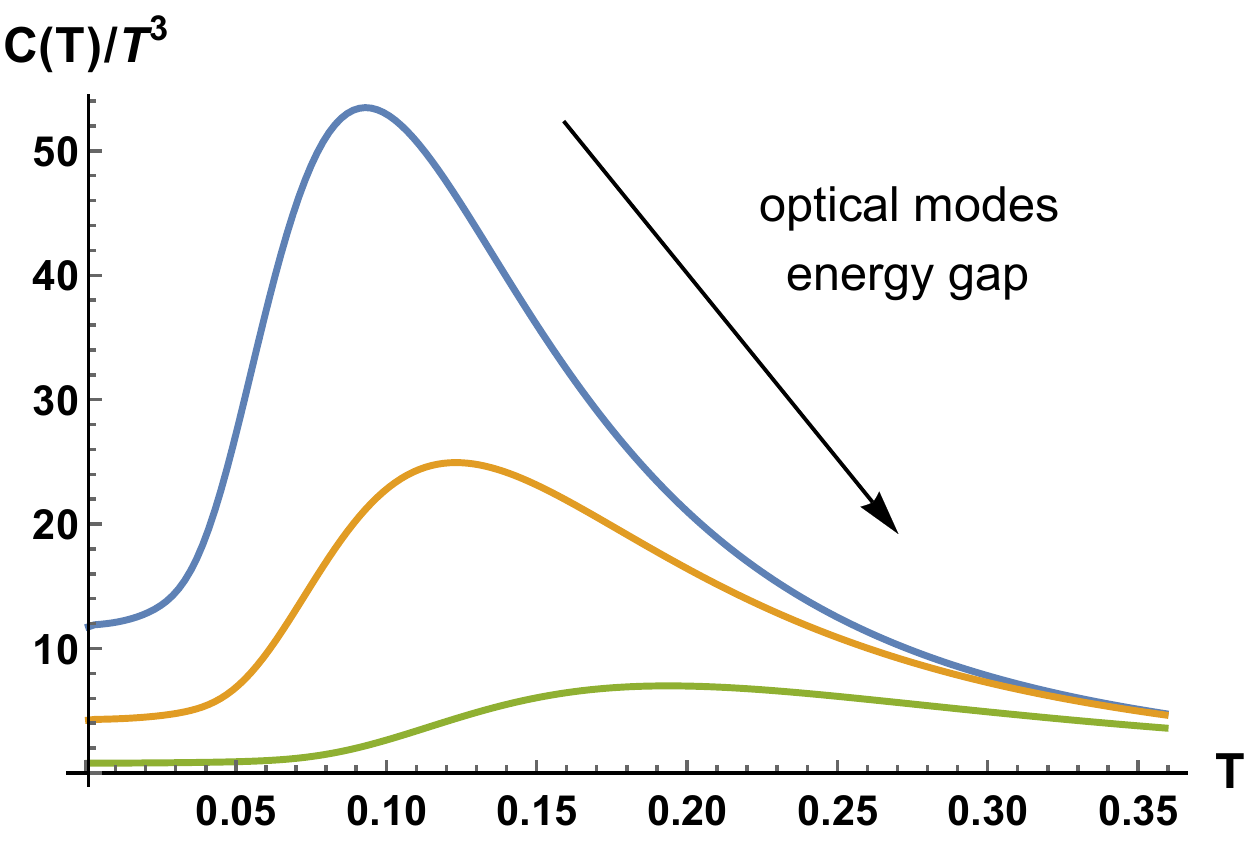}
\centering

\vspace{0.5cm}

\includegraphics[width=8.5cm]{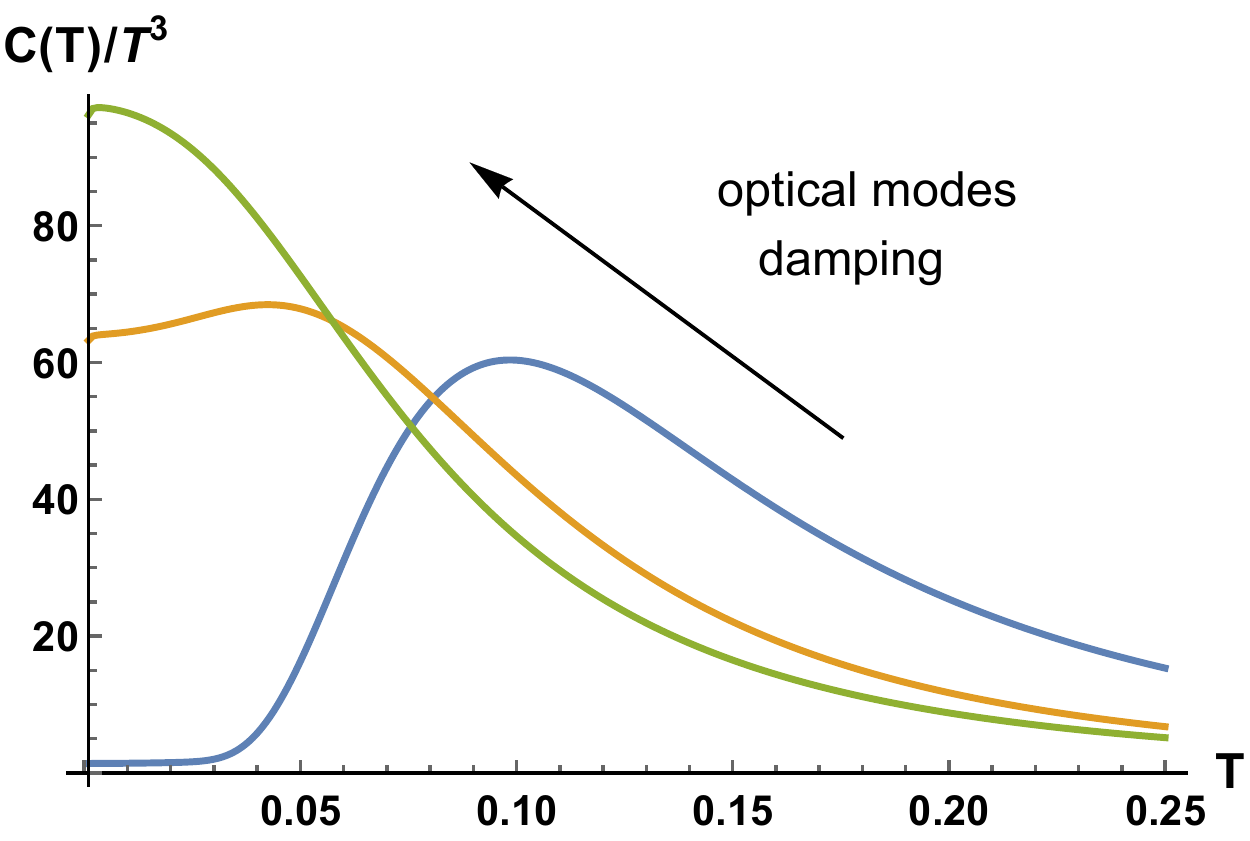}
\caption{\textbf{Top panel:} Optical phonon contribution (only) to the specific heat calculated with Eq.\eqref{eq9} on the basis of the VDOS given by Eq.\eqref{eq8} upon varying the optical energy gap $\omega_{0}$. In the calculation we took $\hbar=1$ and $k_{B}=1$. \textbf{Bottom panel:} The same quantity plotted upon varying the magnitude $D_0$ of the optical phonon damping coefficient $D_{O}$. }
\label{fig5}
\end{figure}\vspace{0.5cm}
In Fig. 4 bottom panel we studied the same model with a different set of parameters, in which the shear elastic constant has a larger value, whereas the damping for the optical phonon is larger. The latter effect shifts the boson peak to lower frequencies, as shown in Fig.\ref{fig4}, and in turn the red-shifted boson peak produces an upturn in the specific heat as $T \rightarrow 0$. This upturn is highly reminiscent of the two-level systems (TLS) tunneling mechanism which is usually invoked to explain this upturn of the specific heat at low $T$. However, the TLS mechanism assumes a random distribution of TLS, and hence it is unlikely to explain this upturn which has been observed experimentally in ordered crystals in ~\cite{Moratalla}. Instead, the mechanism proposed here for the upturn does not rely on any disorder assumption, and is uniquely provided by the red-shift of the boson peak caused by the strongly damped optical phonon. 

In Fig. 5 the optical contribution (only) to the specific heat is plotted. In particular, in Fig. 5 (top panel) the optical contribution is plotted upon varying the optical energy gap $\omega_{0}$, and the peak is shown to become larger and shifted towards low-T as the optical energy gap is reduced. 

In Fig. 5 (bottom panel), the same optical contribution is plotted, this time upon varying the damping coefficient of the optical phonon. It is seen that the peak in the specific heat is strongly shifted towards $T=0$ as the damping of the optical phonon is increased. In particular, for strong damping the optical contribution is shown to produce an upturn in the specific heat at low-T which closely resembles the TLS-like behaviour that has been observed many times in glasses.

\subsection{Effect of the piling up of optical phonons}
Next, we study the effect of having more optical phonons piled up at low energy, as schematically depicted in Fig. 1. We use the full Eq. (8) to evaluate the VDOS, and we vary $N_{O}$ from 1 to 4. In Fig. 6 we show the effect of increasing the number of optical phonons $N_{O}$ piled up at low energy on the VDOS, for a fixed set of all the other parameters. It is clear that the boson peak increases significantly upon increasing the number of optical phonons, although its position is only slightly blueshifted as $N_{O}$ is increased. Upon increasing the number of optical modes, furthermore, the upturn at low $\omega$ appears to increase.

\begin{figure}
\centering
\includegraphics[width=8.5cm]{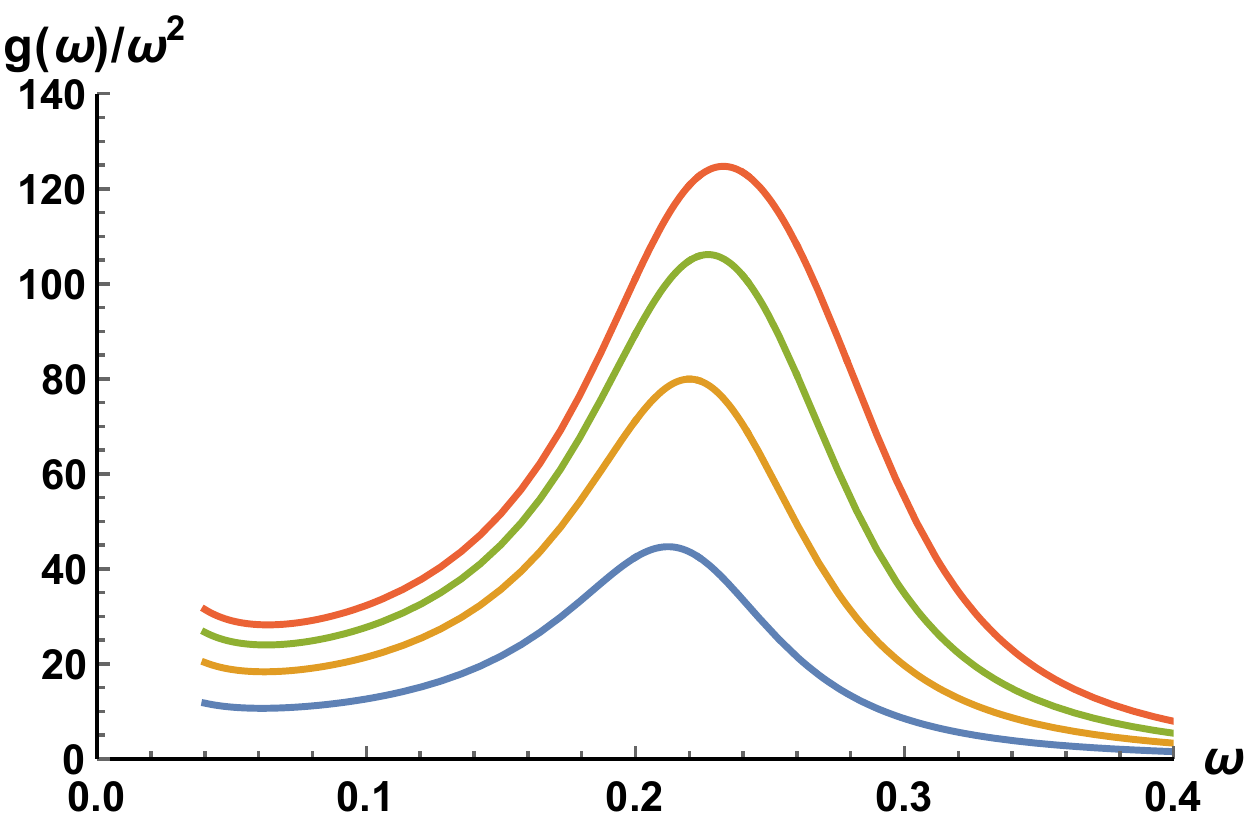}
\caption{VDOS computed upon varying the number of optical phonons, $N_{O}$, from 1 (bottom) to 4 (top curve). The phonon levels (as depicted schematically in Fig.\ref{fig1}) are equally spaced with a fixed interval equal to $\Delta\omega=0.02$, in $\omega_{0}$. }
\label{fig7}
\end{figure}\vspace{0.5cm}

\begin{figure}
\centering
\includegraphics[width=8.5cm]{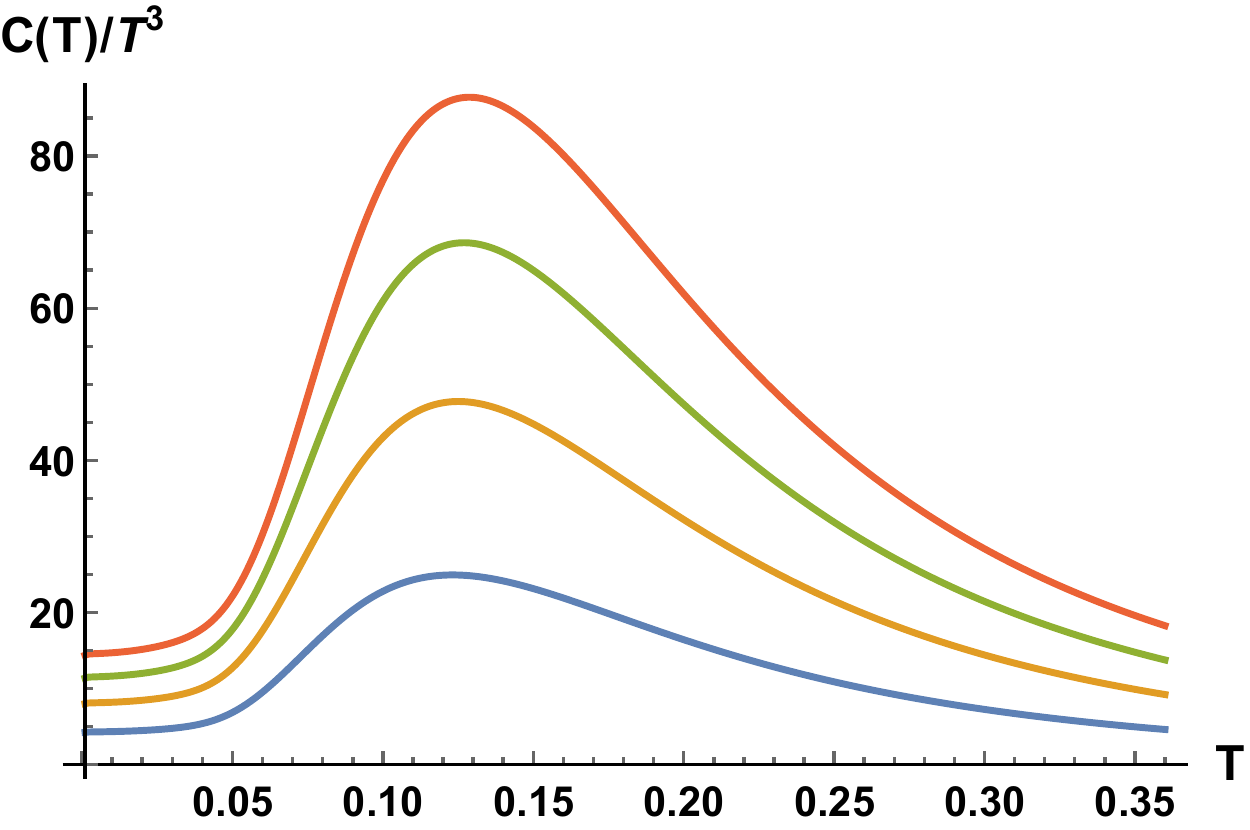}
\caption{Specific heat computed using the VDOS of Fig.\ref{fig7}, i.e. upon varying the number of optical phonons, $N_{O}$, from 1 (bottom) to 4 (top curve). The phonon levels (as depicted schematically in Fig.\ref{fig1}) are equally spaced with a fixed interval equal to $\Delta\omega=0.02$, in $\omega_{0}$. }
\label{fig8}
\end{figure}

In Fig. 7 we plot the effect of varying $N_{O}$ within the same range, and using the same parameters as in Fig. 6, on the specific heat. It is clear that an increase in the number of optical phonons piled up at low energy produces a significant increase of the peak in the specific heat, while the peak position is only slightly shifted towards higher T.

\subsection{Prediction of linear-in-T and TLS-like features in crystals with strongly damped low-energy optical phonons}\label{TLS}
Finally, we look more closely at what happens when the optical phonon is strongly damped. In Fig. 8 we show the VDOS (unnormalized) of the strongly damped optical phonon, which displays a constant plateau at relatively low frequencies, for strong enough damping. In particular, we can notice that at small values of the damping no plateau $g(\omega)\sim const$ appears. On the contrary for large values of the diffusive-like damping prefactor $D_{0}$, a very robust and frequency independent intermediate regime arises. Moreover, the larger the damping the closer the onset of this plateau to zero frequency. In Fig. 9 we show the total (acoustic plus optical) specific heat, which displays a strong upturn towards $T \rightarrow 0$. 

In Fig. 10, we plot the optical contribution only to the specific heat. In Fig. 10 (bottom panel)  we plot the same data but this time in log-log plot. It is clear that this contribution to the specific heat goes exactly like $\sim T$ at low-T, which is the same behaviour observed in glasses and traditionally attributed to the Two-Level-System (TLS) mechanism, based on quantum tunnelling between randomly distributed double wells \cite{Phillips1,Phillips2,Anderson}. 
However, recently, the same upturn and characteristic behavior shown here has been detected in perfectly ordered crystals of organic molecules, e.g. in ~\cite{Moratalla}, where low-energy optical phonons have been shown (by means of DFT calculations) to play a prominent role in the lattice dynamics and in the VDOS. 
Hence our prediction of a TLS-like behaviour in crystals with low-energy and strongly damped phonons appears verified in comparison with those observations.

\begin{figure}[hbtp]
\centering
\includegraphics[width=8.5cm]{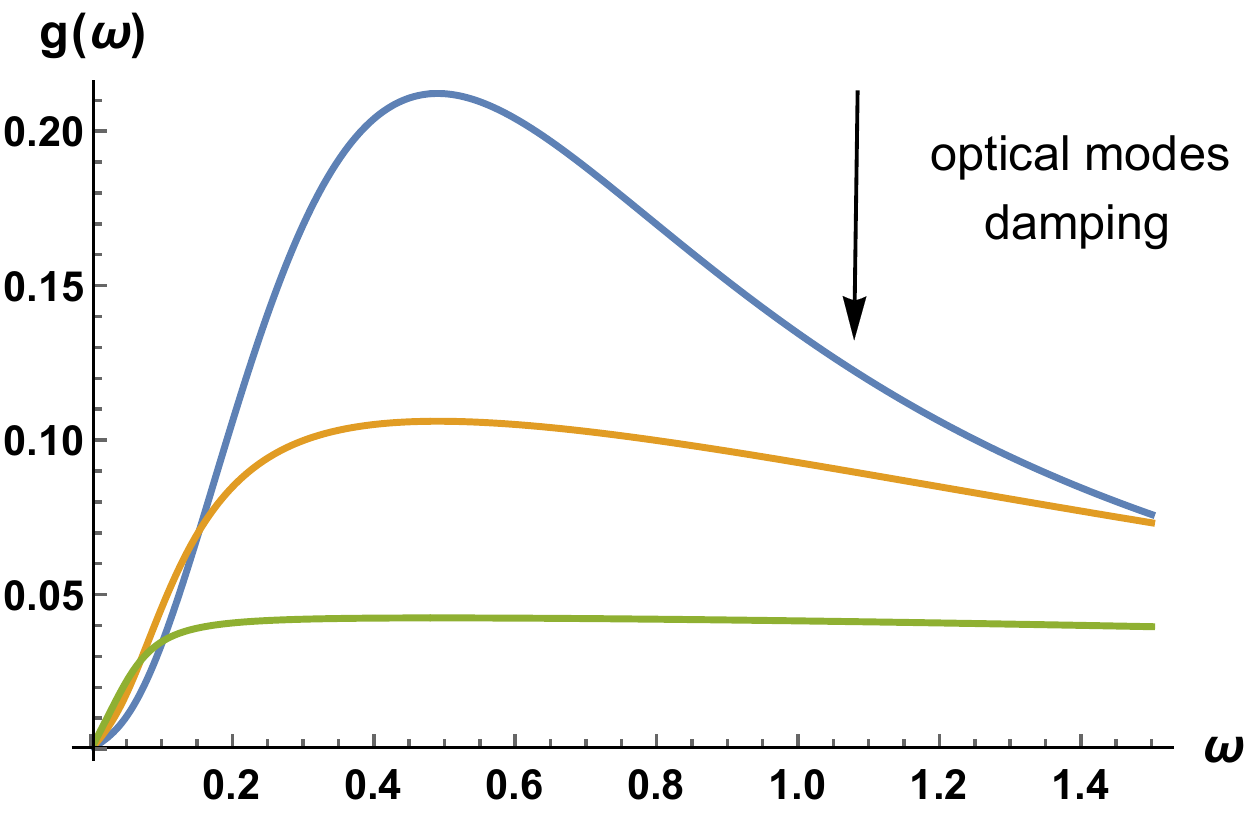}
\caption{VDOS (unnormalized) for a strongly damped optical phonon. A plateau in $\omega$ is reached for strong enough damping, which is typical of diffusons. }
\label{fig9}
\end{figure}

\begin{figure}[h]
\centering
\includegraphics[width=8.5cm]{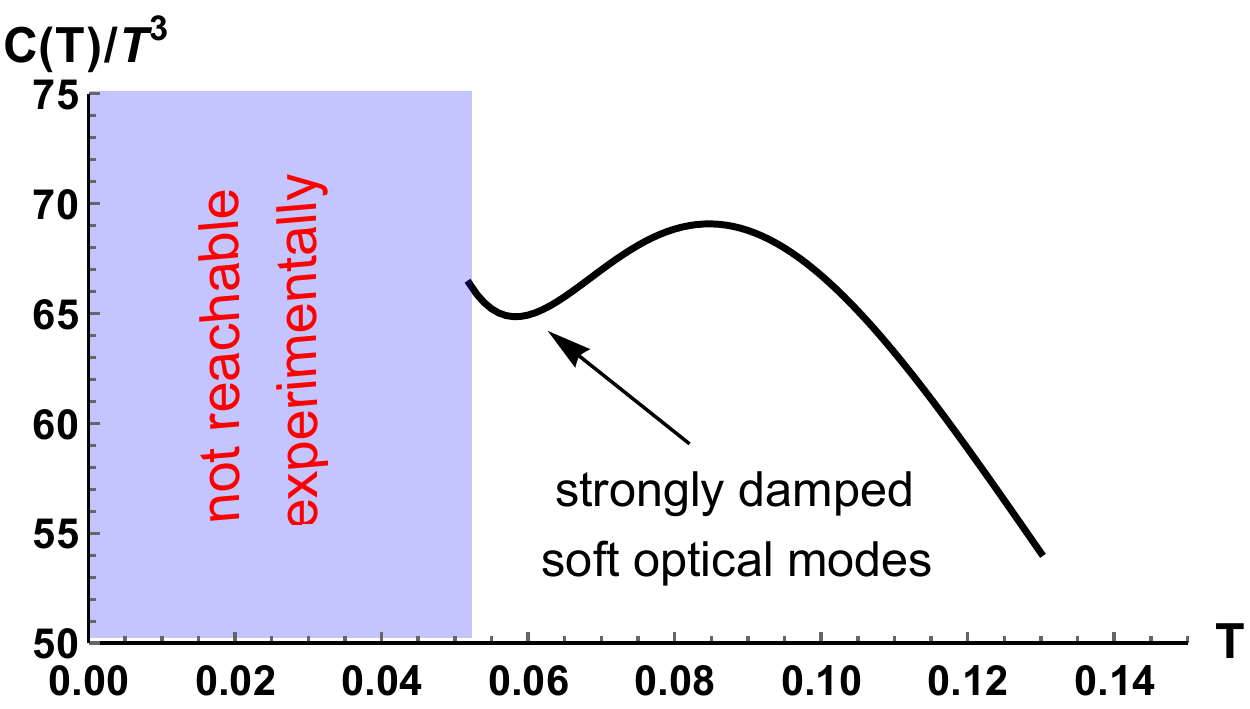}
\caption{Total specific heat computed considering a soft (low-energy) optical phonon with a large diffusive-like damping coefficient. The strong damping produces an upturn at low temperature similar to what is predicted by the TLS model. }
\label{fig6}
\end{figure}

\section{Link with Gardner physics in disordered solids}
Our model presents a description of phonons in a lattice within an effective field theory strategy, which means that we do not explicitly model the microstructure of the solid. The latter may be a highly ordered crystal with a complex unit cell formed by several different atoms leading to many optical modes. Also, it may be a crystal lattice with defects (inclusions, vacancies, polydispersity etc) which vibrate within the lattice matrix and may couple to the acoustic field of the matrix. 

In the case of slightly disordered crystals, an important connection could be drawn between our model and the Gardner transition, originally proposed for spin glasses~\cite{Gardner} and then re-discovered in numerical hard-sphere glassy systems~\cite{Zamponi,Berthier}, and, recently, also in slightly disordered crystals~\cite{Charbonneau}. At the Gardner transition, the local minima which form a meta-basin in the energy landscape of a glassy solid develop a fractal hierarchy of sub-minima within a local minimum. This phenomenon is deeply connected with lifting the degeneracy of particle contact forces, which happens typically upon increasing the packing fraction or the pressure, e.g.  in hard-spheres and possibly deep in the glass state, well below the glass transition. The Gardner phase is a marginally stable state with many similarities to jammed packings in terms of degeneracy of the local contact network. Very similar physics has been recently found in Ref.~\cite{Charbonneau} while studying slightly disordered crystals formed from polydisperse spheres. The polydispersity lifts the degeneracy of vibrational modes and introduces a variety of optical modes with different energy, resulting in a VDOS similar to that of jammed packings at the jamming (isostatic) point where the VDOS remains flat down to zero frequency. This is a situation where basically the boson peak effectively occurs at zero frequency. 

Our model presents a physics very similar to what has been found in the polydisperse crystals of~\cite{Charbonneau}:
\begin{itemize}
\item[(I)] Increasing the number of optical modes towards low frequency similarly lifts the degeneracy of the modes and introduces a proliferation of soft modes, as in the Gardner state. This is associated with the drifting of the boson peak toward zero-frequency, hence toward a marginally-stable state, again like the Gardner state.
Furthermore, also in highly-ordered crystals such as halomethanes~\cite{Moratalla}, the optical modes, which contribute to the boson peak according to the mechanism presented in our model, are highly degenerate due to different rotation-vibration couplings.
\item[(II)] If the model presented above were to describe optical modes arising from randomly-distributed defects, a coupling between defects and acoustic field can be established, as shown in Ref.~\cite{Baggioli_thermo}, with a tunable parameter which gives the concentration of defects. Upon increasing the concentration of defects, the boson peak frequency drifts toward zero frequency as shown in ~\cite{Baggioli_thermo}. Assuming that the characteristic wavevector goes like a linear dispersion relation (which is certainly valid just below the boson peak), this implies that the characteristic length scale of the disorder grows and diverges~\cite{Berthier_PNAS} as the boson peak drifts toward zero frequency, i.e. upon increasing the concentration of defects. Hence there might be a critical defect concentration below which the system transitions into a marginally-stable state similar to the Gardner state.
\item[(III)] The proliferation of soft modes associated with low-lying optical phonons and the associated boson peak causes a breakdown of standard elasticity as observed for the Gardner phase in Ref.~\cite{Biroli}. In particular, this becomes apparent in the development of memory effects in the elastic modulus, which  becomes non-Markovian and history-dependent. The link between soft modes in the boson peak and non-Markovianity has been studied within the nonaffine elasticity framework recently in the case of metallic glasses~\cite{Wang}.
\end{itemize}
It will be very interesting, in future work, to study how these soft optical modes relate to motions involved in the $\beta$ relaxation of glasses  (a connection that has been recently studied in ~\cite{Cui}), since those are responsible for the jump from one local minimum in the energy landscape to another one nearby, which defines the $\beta$  relaxation process.

\section{Conclusions}

The main predictions and outcomes of the presented theoretical framework are as follows.

(i) The optical contribution to the reduced vibrational density of states (VDOS) $g(\omega)/\omega^2$ gives rise to a boson peak if the optical phonons are at low enough energy (i.e. comparable to acoustic mode energy at the Brillouin zone boundary).

(ii) The contribution of the optical modes is, instead, completely irrelevant if the optical phonons lie at energies much larger compared to the energy of the acoustic phonons at the Brillouin zone boundary. 

(iii) We verified, that the boson peak due to the low-energy optical phonons grows upon increasing the number of low-energy optical phonons, with dispersion relation given by Eq.\eqref{disp}.

(iv) The boson peak is accompanied by an upturn at $\omega \rightarrow 0$, which is given by the optical contribution, and which has been found experimentally in certain systems where low-energy optical modes are active~\cite{Suekuni,Mori1,Mori2}. 

(v) The resulting specific heat presents a boson peak with contributions from both optical and acoustic phonons, where the optical contribution may be the dominant one for certain choices of the parameters. The upturn of the specific heat at low-$T$, for strongly damped optical phonons, is entirely controlled by the low-energy optical phonons, and is closely related to the upturn in the normalized VDOS, which is also controlled by the optical phonons.

(vi) For strongly damped low-energy optical phonons, the linear-in-$T$ anomaly in the specific heat is reproduced in perfectly ordered crystals.

\begin{figure}
\includegraphics[width=8.5cm]{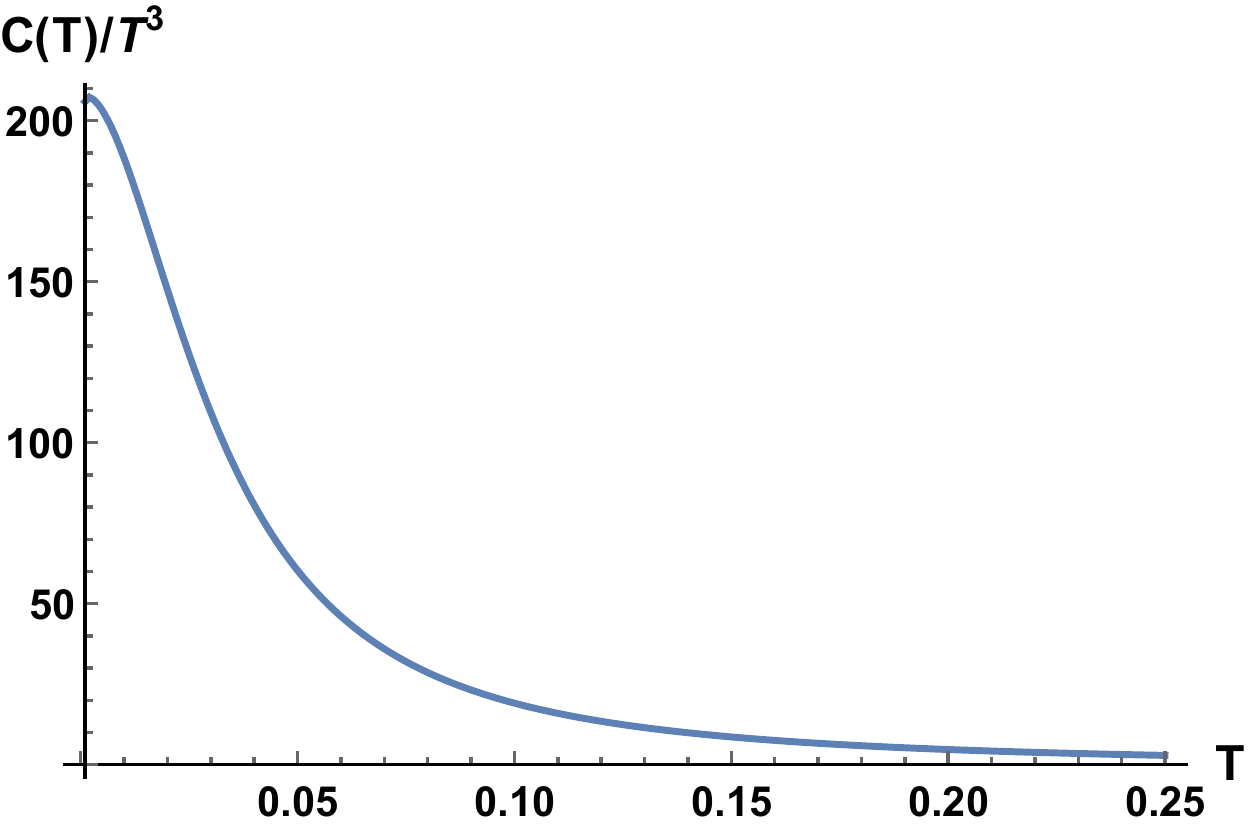}
\centering

\vspace{0.5cm}

\includegraphics[width=8.5cm]{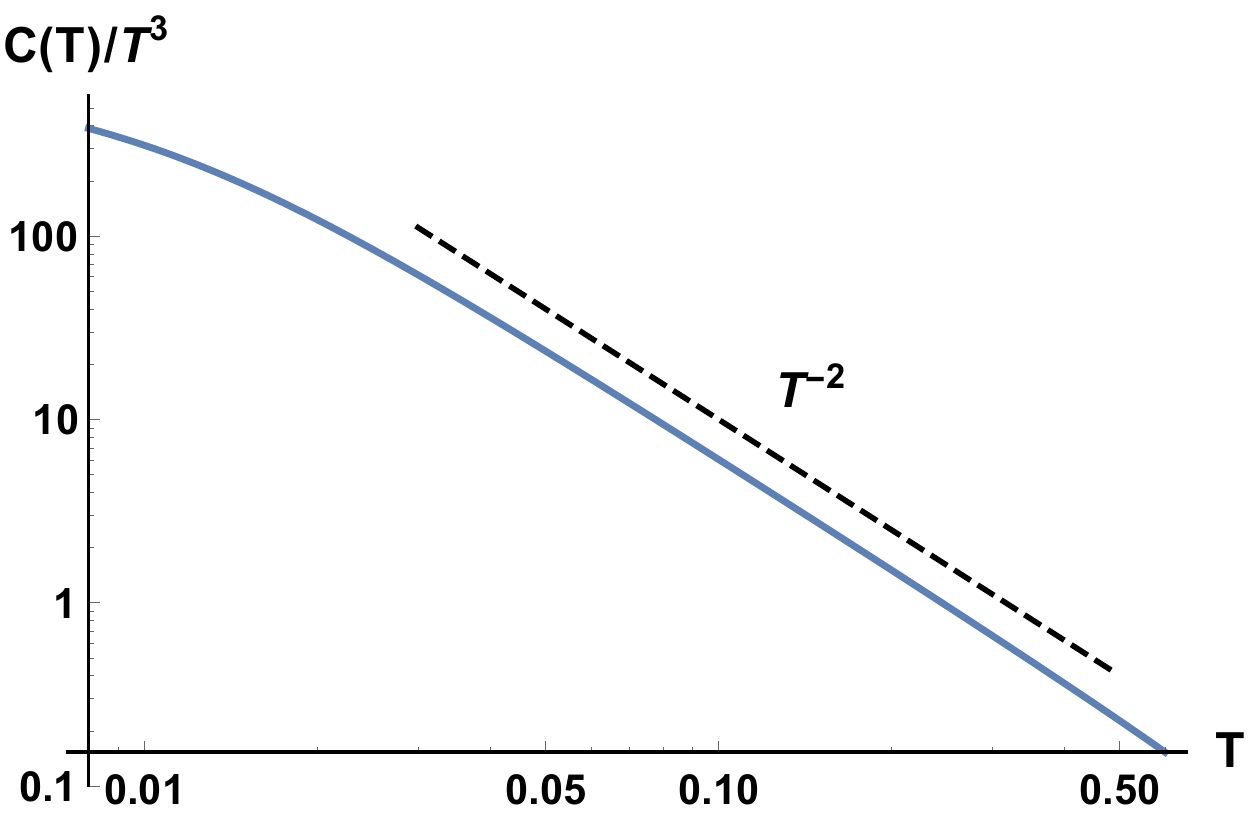}
\caption{\textbf{Top panel:} Optical phonon contribution (only) to the specific heat calculated with the same strongly damped optical phonon the VDOS of which is plotted in Fig.\ref{fig9} . In the calculation we took $\hbar=1$ and $k_{B}=1$. \textbf{Bottom panel:} The same quantity in log-log plot in order to highlight the TLS-like behaviour at low T with the characteristic $C(T)\sim T$ glassy-like trend. }
\label{fig10}
\end{figure}
We also note that the low-energy optical phonons induce a scaling of the specific heat at low-$T$ which approximately goes like $C(T)\sim T$. While it is accepted that the specific heat in glasses goes linearly in $T$ at low $T$, similar non-Debye glassy anomalies have been recently observed in crystals~\cite{Monaco,Tamarit}. This finding can therefore help to shed new light on the origin of these glassy-like anomalies in the specific heat of crystals, where disordered-based TLS arguments are not applicable.

In conclusion, the above theoretical calculation shows that low-energy optical phonons in ordered crystals are responsible for glassy-like anomalies in the vibrational density of states (the so-called boson peak) and in the low-T specific heat. Also, the model shows that, for low-lying optical modes, the normalized VDOS may no longer be flat at $\omega \rightarrow 0$ but may exhibit an upturn, as shown recently in some systems~\cite{Suekuni,Mori1,Mori2}, including thermoelectric crystals~\cite{Suekuni} where low-energy optical modes are important.

The present findings have several implications for various types of solids. For example, in semi-crystalline amorphous superconductors, it is known that the excess of vibrational modes (above the Debye level) gives a large contribution to the Eliashberg function and hence to the electron-phonon coupling, thus enhancing the critical temperature $T_{c}$~\cite{Buckel,Bergmann,Somayazulu}. This effect is proportional to the amount of disorder but vanishes as the material becomes completely amorphous. The finding reported here, that a strong excess of vibrational modes can be achieved thanks to low-energy optical modes in crystals, opens up a new perspective to quantitatively rationalize a large amount of experimental data on conventional superconductors where the variation of $T_{c}$ is largely empirical. Furthermore, it may provide design principles for a new class of high-$T_{c}$ superconductors, by combining light polarons~\cite{Berciu} with the boson peak in perfectly ordered materials.

\begin{acknowledgments}
We would like to thank J.-L. Tamarit for inspiring discussions and for motivating us to study this problem.
We thank M. Dove, M.-A. Ramos, A.Cano, A. Krivchikov, K. Samwer, T. Mori,  and K. Trachenko for several discussions and helfpul comments. We thank M.-A. Ramos for a critical reading of the manuscript. We acknowledge the Thomas Young Centre and the TYC Soiree where this work started.
MB acknowledges the support of the Spanish MINECO’s “Centro de Excelencia Severo Ochoa” Programme under grant SEV-2012-0249.
MB is supported in part by the Advanced ERC grant SM-grav No 669288.  
\end{acknowledgments}

\bibliographystyle{unsrt}

\end{document}